# Complex quantum hydrodynamics with teleportation

Roumen Tsekov
Department of Physical Chemistry, University of Sofia, 1164 Sofia, Bulgaria

It is shown how the Schrödinger equation can be transformed to a complex quantum Navier-Stokes equation with imaginary dilatational viscosity. The self-diffusion in quantum gases is described by this complex quantum hydrodynamics and the quantum Marangoni effect is introduced. The density matrix operator and Wigner function equations, corresponding to quantum hydrodynamics, are described. Finally, the quantum teleportation in time is also discussed via a non-relativistic version of the Majorana equation.

The quantum mechanics was one of the main motors of the scientific revolution in the previous century. Many important quantum problems were solved and found their practical usage. At the beginning of the new century the scientists started desperately to look for novel quantum ideas, mainly related to reinterpretations;[1] at present there are about 20 interpretations of quantum mechanics. At the time when the latter was born, Madelung[2] has proposed a simple analogy between the Schrödinger equation and hydrodynamics. Later Bohm[3] has developed a hidden variable theory, which has grown nowadays to a modern quantum concept.[4] Recently, the Bohmian mechanics was extended to the complex space.[5] The present paper aims to explore a complex version of the Madelung quantum hydrodynamics.

According to nonrelativistic quantum mechanics, the evolution of the wave function $\psi$ of a particle with mass $m$ in vacuum obeys the Schrödinger equation

$$i\hbar \partial_t \psi = \hat{H} \psi \tag{1}$$

Here $\hat{H} \equiv \hat{p}^2 / 2m + U$ is the particle Hamiltonian, where $\hat{p} \equiv -i\hbar \nabla$ and $U$ are the momentum operator and an external potential, respectively. The complex wave function can be generally expressed by a complex action $S$ via the eikonal representation

$$\psi = \exp(iS / \hbar) \tag{2}$$

Introducing this expression in Eq. (1) yields, after simple rearrangements, a complex Hamilton-Jacobi equation[6,7]

$$\partial_t S + (\nabla S)^2 / 2m + U = \mathcal{D} \nabla^2 S \tag{3}$$

As is seen, the quantum effect appears via a diffusive process for the complex action $S$ with an imaginary diffusion constant $\mathcal{D} \equiv i\hbar/2m$. Note that there is no quantum potential in Eq. (3), in contrast to the de Broglie-Bohm theory.[3] If a complex hydrodynamic velocity is introduced via the expression $W \equiv \nabla S/m$, Eq. (3) transforms, after application of a nabla operator $\nabla$, into the Navier-Stokes equation from the classical hydrodynamics

$$\partial_t W + W \cdot \nabla W = -\nabla U/m + \nabla(\xi \nabla \cdot W) \tag{4}$$

Pressure is absent in Eq. (4), since the particle is moving in vacuum, and because the latter is a non-dissipative environment its universal quantum dilatational viscosity $\xi = \mathcal{D}$ is purely imaginary. Note that at high velocity the second nonlinear term here can cause complex quantum turbulence.[8] Equation (4) allows use of the hydrodynamic formalism to solve quantum problems without Hamiltonians. Let us try, for instance, to describe dissipative quantum systems. The rigorous derivation of the quantum gas dynamics follows the quantum BBGKY hierarchy but the final results are too complicated due to mathematical complexity.[9,10,11] For this reason, we will focus on the motion of a single quantum particle in a dilute gas of identical particles. In this case the dilatational viscosity in the complex Navier-Stokes equation (4) will possesses also a real part, being the classical self-diffusion coefficient $D$ in the gas,

$$\xi = D + \mathcal{D} \tag{5}$$

Introducing the mass density via the standard expression $\rho \equiv m\bar{\psi}\psi$, the complex hydrodynamic velocity $W = V - \mathcal{D}\nabla \ln \rho$ splits into real and imaginary parts. The real part $V$ represents the usual hydrodynamic velocity, while the imaginary part is the so-called quantum osmotic velocity.[12] Substituting this representation in Eq. (4), accomplished by Eq. (5), leads to the following equation corresponding to the imaginary part

$$\partial_t \rho + \nabla \cdot (\rho V) = D\rho \nabla^2 \ln \rho = \nabla \cdot (D\nabla \rho) \tag{6}$$

In the derivation of the last expression it is employed that for gases $D\rho$ is constant. Equation (6) is the classical convective diffusion equation. The real part of Eq. (4) yields a real quantum Navier-Stokes equation[2,13,14]

$$\partial_t V + V \cdot \nabla V = -\nabla(U+Q)/m + \nabla(D\nabla \cdot V) \tag{7}$$

where the quantum effects are included in the Bohm quantum potential $Q \equiv -\hbar^2 \nabla^2 \rho^{1/2}/2m\rho^{1/2}$. If one considers the case of no external field ($U=0$) and neglects the two inertial terms on the left hand-side of Eq. (7), it acquires after integration the form $D\nabla \cdot V = Q/m$. Multiplying this by the mass density and using the constantans of the product $D\rho$ yields

$$\nabla \cdot (D\rho V) = \rho Q/m \approx \mathcal{D}^2 \nabla^2 \rho \qquad (8)$$

where the last approximation holds for relatively flat mass distribution. Integrating once again Eq. (8) provides the real part of the complex hydrodynamic velocity

$$V = (\mathcal{D}^2/D)\nabla \ln \rho \qquad (9)$$

which possesses osmotic structure as well. Thus, the ratio of the real and imaginary components scales with ratio between the universal quantum and classical diffusion coefficients.

Introducing the velocity from Eq. (9) in the convective diffusion equation (6) results in an ordinary diffusion equation without convection

$$\partial_t \rho = \nabla \cdot (\mathcal{D}\nabla \rho) \qquad (10)$$

where the effective diffusion coefficient $\mathcal{D} \equiv D - \mathcal{D}^2/D$ is a sum of the classical and quantum diffusion constants.[15,16,17] It possesses a minimum in respect to the classical diffusion constant at $D = \hbar/2m$, which reflects the Heisenberg uncertainty principle. Since $D$ decreases with a density increase, the quantum effect will be emphasized in dense gasses at low temperature, e.g. in the neutron stars, for instance. Similar to the classical case, the quantum hydrodynamics needs necessary boundary conditions reflecting the properties of the fluid interfaces.[18] Thus, one should take into account the adsorption of quantum particles, their surface diffusion and convection, etc. An interesting phenomenon here is the quantum Marangoni effect. After Gibbs, the adsorption of particles decreases the surface tension $\sigma$. Due to the hydrodynamic flow in the bulk the adsorption could be non-homogeneously distributed on the surface, e.g. surface plasmons, which induces a gradient of the surface tension. According to hydrodynamics the latter is compensated by the interfacial value of the bulk stress tensor, i.e. $\partial_x \sigma = \rho D \partial_x V_z$. Using that $\rho D$ is constant one can integrate this equation to obtain

$$\Delta \sigma = \rho D V_z = \mathcal{D}^2 \partial_z \rho \qquad (11)$$

where the last expression is obtained by the use of Eq. (9). Therefore, the non-uniform distribution of quantum particles in the bulk will cause also changed of the surface tension solely due to quantum effects. Moreover, Eq. (11) defines a new quantum screening parameter $\partial_\rho \sigma/\mathcal{D}^2$.

The quantum Marangoni number is defined then by the product of this quantum screening reciprocal length and the characteristic length of the system geometry.

It is interesting to explore how the Schrödinger equation changes by the effect of an environment. In this case the complex Hamilton-Jacobi equation reads

$$\partial_t S + (\nabla S)^2 / 2m + U = \xi \nabla^2 S \tag{12}$$

where the dilatational viscosity possesses real and imaginary components in accordance to Eq. (5). By the use of the wave function definition from Eq. (2), Eq. (12) can be easily transformed to the following Schrödinger equation

$$i\hbar \partial_t \psi = \hat{H}\psi + i\hbar D_\psi \nabla^2 \ln \psi = \hat{H}\psi + i\hbar \nabla \cdot (D_{\bar{\psi}} \nabla \psi)/\bar{\psi} \tag{13}$$

which is a particular example of the Doebner-Goldin equations.[19] The last diffusive term is purely entropic since its average value is zero, i.e. it gives no change in the system energy. For relatively flat wave functions Eq. (13) can be further linearized to

$$\partial_t \psi = -i\hat{H}\psi/\hbar + D\nabla^2 \psi \tag{14}$$

This diffusive Schrödinger equation is a mean field approximation of the quantum state diffusion theory.[20,21] The latter extends the Schrödinger equation to a stochastic differential equation by adding a Wiener process to describe the effect of an environment. In Eq. (14) this white noise is replaced by its macroscopic image, i.e. the classical diffusion. In the case of a free quantum particle Eq. (14) reduces to a diffusion equation with a complex diffusion coefficient

$$\partial_t \psi = (\mathcal{D} + D)\nabla^2 \psi = \xi \nabla^2 \psi \tag{15}$$

The solution of Eq. (15) for a constant dilatational viscosity is a damped plane wave.

According to the quantum mechanics the most complete description of a quantum system is given in terms of the wave function. For this reason, the classical notion of phase space probability density is replaced by the density matrix operator $\hat{\rho}$. In the case of the Schrödinger equation (1) the density matrix operator evolves in time via the von Neumann equation

$$\partial_t \hat{\rho} = -i[\hat{H}, \hat{\rho}]/\hbar \tag{16}$$

The formal solution of Eq. (16) reads $\hat{\rho}(t) = \exp(-i\hat{H}t/\hbar)\hat{\rho}(0)\exp(i\hat{H}t/\hbar)$, which acquires the following form in the energy basis

$$\hat{\rho} = \sum\sum \exp[i(E_n - E_k)t/\hbar]|E_k\rangle\langle E_k|\hat{\rho}(0)|E_n\rangle\langle E_n| \tag{17}$$

where $\{E_n\}$ are the energy eigenvalues of the Hamiltonian $\hat{H}$. As is seen from Eq. (17) the density matrix possesses non-diagonal elements, while the equilibrium density matrix, following from the quantum statistical physics, is diagonal $\hat{\rho}_{eq} = \sum p_k |E_k\rangle\langle E_k|$ with $p_k$ being the probability for occupation of the state $|E_k\rangle$. To explain this matrix reduction the modern theory of decoherence shows that the effect of an environment is to destroy the non-adaptive non-diagonal elements, a concept known as the quantum Darwinism.[22] In this respect, one can derive from Eq. (14) a new master equation

$$\partial_t \hat{\rho} = -i[\hat{H}, \hat{\rho}]/\hbar - D\{\hat{p}^2, \hat{\rho}\}/\hbar^2 \tag{18}$$

which describes a new type of decoherence. Here the brackets [ , ] and { , } denote commutator and anticommutator, respectively. The last diffusive term is a particular example of a part of the Lindblad super-operator,[23] ensuring complete positivity of the solutions of the master equation. Note that the usual Ohmic frictional and thermal terms are not present in Eq. (18), respectively in Eq. (7), since we describe quantum self-diffusion in vacuum.

The problem of the quantum Darwinism is that diagonalization of the density matrix occurs also in isolated systems. Hence, the environment is not absolutely essential for decoherence. As is seen from Eq. (17), the solution of Eq. (16) is a periodic function of time, thus reflecting the Poincare cycles as well. In fact, the evolution never stops and the stationary equilibrium distribution is an idealization, when the fluctuations are somehow omitted. It is believed, however, that one could eliminate the effect of the persistent fluctuations by averaging in time. This so-called ergodic theorem allows us to express the equilibrium density matrix operator from the exact solution of Eq. (16) in the form

$$\hat{\rho}_{eq} = \lim_{\tau\to\infty} \frac{1}{\tau} \int_0^\tau \hat{\rho} dt \tag{19}$$

According to this definition the equilibrium distribution is the most frequently occupied one. Introducing the density matrix operator from Eq. (17) and performing the integration on time leads straightforward to

$$\hat{\rho}_{eq} = \sum \delta_{E_n E_k} |E_k\rangle\langle E_k|\hat{\rho}(0)|E_n\rangle\langle E_n| = \sum |E_k\rangle\langle E_k|\hat{\rho}(0)|E_k\rangle\langle E_k| \qquad (20)$$

where the last expression presumes a non-degenerated energy spectrum of the system. Identifying the probability density $p_k = \langle E_k|\hat{\rho}(0)|E_k\rangle = \delta_{EE_k}$ of the micro-canonical ensemble, Eq. (20) reduces to the diagonal expression known from the equilibrium quantum statistical physics. The consideration above shows that decoherence in isolated systems is caused by the quantum evolution itself and the averaging in time leads to mutual cancelation of the non-diagonal fluctuating elements. It is expected that this self-decoherence mechanism takes place in open systems as well, thus assisting decoherence caused by the environment.

An alternative presentation of quantum mechanics is possible via the Wigner function

$$f(p,r,t) \equiv \int_{-\infty}^{\infty} \exp(ipq)\bar{\psi}(r - \hbar q/2, t)\psi(r + \hbar q/2, t) dq/2\pi \qquad (21)$$

which is a quasi-distribution function in the forbidden phase space of the quantum system. If we accomplish Eq. (18) by the complete form of the decoherence Lindblad super-operator, the corresponding Wigner-Liouville equation reads

$$\partial_t f = \{H, f\}_{MB} + D\nabla^2 f \qquad (22)$$

where $\{,\}_{MB}$ denotes the Moyal brackets. As is seen, a classical diffusion operator appears in Eq. (22) by acting in the coordinate space.[15,24,25] Note that this diffusional term should remain also in the classical limit of Eq. (22), i.e. in the classical Klein-Kramers equation.[26] It is interesting that the whole quantum mechanics consists in Eq. (21). For instance, it follows after integration directly that the probability density in the momentum space

$$g \equiv \int_{-\infty}^{\infty} f(p,r,t) dr = \bar{\phi}\phi \qquad (23)$$

is a product of complex-conjugated wave functions $\phi(p,t) = \int \psi(r,t)\exp(ipr/\hbar)dr$, being the Fourier images of the wave functions in the coordinate space. Moreover, Eq. (21) provides for the mass density and hydrodynamic velocity the well-known expressions

$$\rho \equiv m\int_{-\infty}^{\infty} f(p,r,t)dp = m\bar{\psi}\psi \qquad V \equiv \int_{-\infty}^{\infty} pf(p,r,t)dp/\rho = \mathcal{D}\nabla\ln(\bar{\psi}/\psi) \qquad (24)$$

Introducing these relations in the compulsory continuity equation $\partial_t \rho + \nabla \cdot (\rho V) = 0$, which follows from the law of mass conservation, yields

$$(i\hbar \partial_t \psi + \hbar^2 \nabla^2 \psi / 2m) / \psi = (-i\hbar \partial_t \bar{\psi} + \hbar^2 \nabla^2 \bar{\psi} / 2m) / \bar{\psi} \tag{25}$$

This equation shows that the two complex-conjugated sides are equal to a real function, which is evidently the potential energy $U$. Therefore, we derived the Schrödinger equation (1) only by the use of the Wigner function definition (21) and the laws of mass and energy conservation.

Finally, we would like to explore the applicability of the complex quantum hydrodynamics to the quantum teleportation in time.[27,28,29,30] It is well known that the complex-conjugated wave functions $\psi$ and $\bar{\psi}$ describe evolutions forwards and backwards in time, respectively.[12] Since in the common quantum mechanics the dual spaces are not interacting, the two evolutions do not cross each other. Let us suppose now that there is a way for tunnelling between the dual spaces and as an example we propose here the following Schrödinger equation

$$i\hbar \partial_t \psi = E\psi + \varepsilon \bar{\psi} \tag{26}$$

where $\varepsilon$ is a real coupling constant. Substituting the wave function from Eq. (2) yields

$$\partial_t S_{Re} = -E - \varepsilon \cos(2 S_{Re} / \hbar) \qquad \partial_t S_{Im} = \varepsilon \sin(2 S_{Re} / \hbar) \tag{27}$$

The solution of the first equation $S_{Re} = \hbar \arctan\{\sqrt{E^2 - \varepsilon^2} \tan[\sqrt{E^2 - \varepsilon^2}\, t / \hbar] / (E - \varepsilon)\}$ reduces to the standard expression $S_{Re} = -Et$ in the case of $E \gg \varepsilon$. Substituting this result in the second equation leads to

$$\partial_t S_{Im} = -\varepsilon \sin(2Et / \hbar) \tag{28}$$

One can easily recognize in this equation the superconductive Josephson current,[31] flowing between the dual spaces. A further integration of Eq. (28) leads to $S_{Im} = -\hbar(\varepsilon / 2E) \cos(2Et / \hbar)$ and, thus, the complete wave function from Eq. (2) acquires the form

$$\psi = \exp[-iEt / \hbar + (\varepsilon / 2E) \cos(2Et / \hbar)] \tag{29}$$

As is seen from Eq. (29), the amplitude of the wave function is also fluctuating in time and the corresponding mass density is given by the expression

$$\rho = m\exp[(\varepsilon/E)\cos(2Et/\hbar)] \approx m + m'\cos(2Et/\hbar) \qquad (30)$$

This equation shows that the conservation of matter is correct in average when considered for a long time. It could be, however, temporary violated due to exchange of matter between different time moments. The frequency $2E/\hbar$ of these fluctuations reflects the Heisenberg time-energy uncertainty relation, which is also responsible for existence of virtual particles. Perhaps, the mass of the exchanged virtual particles scales with the mass $m' \equiv m\varepsilon/E$.

In the case of spatially distributed systems the picture above can be extended via the following Schrödinger equation

$$i\hbar \partial_t \psi = \hat{H}\psi + \varepsilon\bar{\psi} \qquad (31)$$

Certainly, in a more advanced treatment $\varepsilon$ should also be replaced here by an operator. Note that Eq. (31) is still a linear differential equation and the superposition principle holds. Applying now Eq. (2) this equation reduces to a new complex Hamilton-Jacobi equation

$$\partial_t S + (\nabla S)^2/2m + U = \mathcal{D}\nabla^2 S - \varepsilon\exp(-2iS_{Re}/\hbar) \approx \mathcal{D}\nabla^2 S - \varepsilon + 2i\varepsilon S_{Re}/\hbar \qquad (32)$$

where the last approximation presumes a small value of the ratio $S_{Re}/\hbar$. Applying a nabla operator to Eq. (32), the latter converts into another complex Navier-Stokes equation

$$\partial_t W + W \cdot \nabla W = -\nabla U/m + \nabla(\mathcal{D}\nabla \cdot W) - \varepsilon V/m\mathcal{D} \qquad (33)$$

The last term describing the time travel is, in fact, the classical Darcy flux through a porous media but with an imaginary quantum permeability $m\mathcal{D}/\varepsilon$. Therefore, the quantum teleportation in time corresponds to a quantum Brownian motion at zero temperature with imaginary friction coefficient. Vice versa, the quantum Brownian motion can be described by Eq. (31) with an imaginary coupling constant $\varepsilon = \mathcal{D}b$, where $b$ is the Brownian particle friction coefficient. In this case Eq. (33) splits into the following two equations

$$\partial_t \rho + \nabla \cdot (\rho V) = 0 \qquad \partial_t V + V \cdot \nabla V = -\nabla(U+Q)/m - bV/m \qquad (34)$$

which are the base of the hydrodynamic description of the quantum diffusion.[32,33] Thus, a complex $\varepsilon = m\mathcal{D}^2\kappa^2 + \mathcal{D}b$ will be able to describe the quantum teleportation through a dissipative environment and the corresponding quantum hydrodynamic equations read

$$\partial_t\rho+\nabla\cdot(\rho V)=\kappa^2\rho\int V dr \qquad \partial_t V+V\cdot\nabla V=-\nabla(U+Q)/m-bV/m \qquad (35)$$

The teleportation current in the first equation looks like a first order chemical reaction with an oscillating in time reaction constant driven by the second quantum Navier-Stokes equation. Note that Eq. (35) is approximate due to the linearization made in Eq. (32). The latter is correct at strong friction, where one can neglect also the first two inertial terms in the second equation to obtain an expression for the hydrodynamic velocity $V=-\nabla(U+Q)/b$. Introducing it in the first of Eq. (35) yields a quantum Smoluchowski equation with teleportation

$$\partial_t\rho=\nabla\cdot[\rho\nabla(U+Q)/b]+\kappa^2\rho(E-Q-U)/b \qquad (36)$$

showing that the energy variations are driving the quantum teleportation. Their mean value is zero, however, which is a necessary condition for the conservation on matter in average. The quantum potential $Q$, being the symbol of the quantum non-locality,[34] is the main promoter either of the quantum diffusion or of the quantum teleportation. Naturally, the effect of the friction is to slowdown the teleportation in time and the equilibrium solution of Eq. (36) obeys the stationary Schrödinger equation $Q+U=E$. Note that in this case both the diffusion and teleportation fluxes vanish. Recently, the friction constant of a quantum particle in a classical gas was evaluated as $b\approx\hbar/\lambda^2$, where $\lambda$ is the mean free path.[17] To satisfy the time-energy Heisenberg uncertainty relation it follows from Eq. (36) that $\kappa\lambda<1$, i.e. the characteristic teleportation length $1/\kappa$ is larger than the mean free path of the medium particles.

In the case of a free quantum Brownian particle ($U=0$) the solution of Eq. (36) is a zero centered Gaussian distribution with position dispersion $\sigma_x^2$ obeying the following equation

$$\sigma_x^2-(2/\kappa^2)\ln(1+\sigma_x^2\kappa^2/2)=\hbar^2\kappa^2 t/4mb \qquad (37)$$

which formally coincides with an equation derived for the quantum Brownian motion.[35] Hence, the teleportation plays effectively a role of temperature $T_T\equiv\hbar^2\kappa^2/8mk_B$ and the transferred momentum equals to $\sigma_p=\hbar\kappa/2$. At short time Eq. (37) reduces to the result $\sigma_x^2=\hbar\sqrt{t/mb}$ for the quantum Brownian motion at zero temperature,[33,35] while at large time the classical Einstein law $\sigma_x^2=2D_T t$ follows but with a quantum diffusion constant $D_T\equiv k_B T_T/b$ for teleportation in a dissipative environment. Furthermore, one can heuristically[35] add to Eq. (36) the effect of the non-zero temperature $T$ of the environment to obtain

$$\partial_t\rho=\nabla\cdot[\rho\nabla(U+Q+k_B T\ln\rho)/b]+\kappa^2\rho(F-k_B T\ln\rho-Q-U)/b \qquad (38)$$

where $F$ is the Helmholtz free energy. Since the energy variations are strongly related to the entropic ones, the temperature will affect the teleportation according to Eq. (38). For instance, in the case of a free quantum Brownian particle the dispersion of the corresponding Gaussian distribution is given by

$$(2/\kappa^2)\ln(1+\sigma_x^2\kappa^2/2) - \lambda_T^2\ln(1+\sigma_x^2/\lambda_T^2) = (2-\kappa^2\lambda_T^2)Dt \tag{39}$$

where $\lambda_T = \hbar/2\sqrt{mk_BT}$ is the thermal de Broglie wavelength and $D = k_BT/b$ is the classical Einstein diffusion constant. Equation (39) contains all the limiting cases discussed before and at large time provides an exponential law. At high temperature $T > T_T$ one can neglect the second term in Eq. (39) to obtain

$$\sigma_x^2\kappa^2/2 = \exp(D\kappa^2 t) - 1 \tag{40}$$

Obviously the teleportation accelerates the thermal diffusion and if $\kappa$ is large Eq. (39) leads to $\sigma_x^2/\lambda_T^2 = \exp(D\kappa^2 t) - 1$. Since $\kappa$ is certainly a quantum parameter, Eq. (38) reduces to the ordinary Smoluchowski equation in the classical limit.

As was demonstrated in the paper, the complex quantum hydrodynamics is a powerful tool for description of dissipative quantum systems. The complex hydrodynamic velocity is also the essence of the scale relativity quantum theories.[36] A new concept of quantum teleportation in time is introduced, which differs substantially to the existing theories of quantum teleportation, based on the Einstein-Podolsky-Rosen paradox.[27] A quantum Smoluchowski equation is derived, which accounts simultaneously for thermal and quantum diffusions as well as for the quantum teleportation. Surprisingly, the effect of the quantum teleportation on a free quantum Brownian particle leads to the classical Einstein law but with a new quantum diffusion constant. Finally, Eq. (26) represents, in fact, a non-relativistic Majorana-like equation.[37] Therefore, the time-travel teleportation occurs via a Josephson superconductive flow of Majorana particles. The Majorana fermions were recently discovered in ferromagnetic iron atomic chains adsorbed on the surface of superconducting lead.[38]